\documentclass[twocolumn,aps,pra,superscriptaddress,showpacs]{revtex4}
\usepackage{graphicx,dcolumn}
\usepackage{amsfonts,amssymb,amsmath,mathbbol}
\usepackage[english]{babel}
\begin{document}

\author{Thomas Wellens}
\affiliation{Institut f\"ur Physik, Albert-Ludwigs-Universit\"at
 Freiburg, Hermann-Herder-Str. 3, D-79104 Freiburg, Germany}
\author{Tobias Geiger}
\affiliation{Institut f\"ur Physik, Albert-Ludwigs-Universit\"at
 Freiburg, Hermann-Herder-Str. 3, D-79104 Freiburg, Germany}
\author{Vyacheslav Shatokhin}
\affiliation{Institut f\"ur Physik, Albert-Ludwigs-Universit\"at
 Freiburg, Hermann-Herder-Str. 3, D-79104 Freiburg, Germany}
\affiliation{B. I. Stepanov Institute of Physics
  NASB, 220072 Minsk, Belarus}
\author{Andreas Buchleitner}
\affiliation{Institut f\"ur Physik, Albert-Ludwigs-Universit\"at
 Freiburg, Hermann-Herder-Str. 3, D-79104 Freiburg, Germany}
\title{Scattering laser light on cold atoms: multiple scattering signals from single-atom responses}
\date{\today}

\begin{abstract}
We deduce the coherent backscattering signal from two distant laser-driven
atoms using single-atom equations. In contrast to the
standard master equation treatment, this new approach is suitable for
the generalization to a large number of atomic scatterers. 
\end{abstract}

\pacs{42.50.Ct, 34.50.Rk, 42.25.Dd}

\maketitle

Cooled atomic gases have been successfully used in
experimental studies of coherent backscattering (CBS) of laser
light \cite{labeyrie99}, and are considered to be a promising medium
for the observation of Anderson localization \cite{anderson58} with
photons. Apart from their fundamental interest, these interference phenomena
in disordered systems are important for practical applications
 such as random lasers
\cite{cao} and quantum memory for light \cite{lukin03}. Inasmuch as
 \emph{multiple} scattering of photons is crucial for
the emergence of weak and strong localization \cite{meso94} in
disordered atomic clouds, while it can play a detrimental role for the
storage of quantum information \cite{datsyuk06}, an appropriate
theory taking into account all relevant effects of atom-light
interactions is in order.

The theory of multiple scattering in
dilute media that consist of a disordered collection of discrete
scatterers relies on the division of the total scattering process
into single scattering events. These are described by a scattering
operator
which is usually assumed to be
identical for all the individual scatterers, and thereby serves as
the fundamental building block for the total multiple scattering
process \cite{tmatrix}. This approach, however, fails in the case of
atomic scatterers, for the following two reasons: first, due to
the saturation of the atomic transition with increasing intensity of
the incident light, the atoms scatter light nonlinearly. The
outgoing field ($E_{\rm out}$) is not proportional to the incoming one
($E_{\rm in}$), and therefore cannot be described by a linear scattering
operator of the form $E_{\rm out}=T E_{\rm in}$. Second, the light
scattered by near-resonant atoms exhibits fluctuations due to the
quantum mechanical coupling of the atoms to the electromagnetic
vacuum. This means that even if the incident laser light
$E_{\rm in}$ is perfectly coherent,
this is not the case for the scattered field $E_{\rm out}$. In
particular, its average intensity differs from the square of the
average field, i.e., $\langle|E_{\rm out}|^2\rangle > |\langle
E_{\rm out}\rangle|^2$. The difference defines the incoherent (or
inelastic) component of the resonance fluorescence intensity.

So far, no satisfying theory
exists for incorporating both these effects into a multiple scattering
approach. Recently \cite{wellens07,wellens09}, a theory for coherent
backscattering by nonlinear, classical scatterers was presented, but
does not take into account any quantum fluctuations due to
inelastic scattering.
A perturbative approach based on the scattering matrix of
two photons was proposed in \cite{wellens04}, but is only valid if
incident light intensity and optical thickness of the atomic
medium are small. On the other
hand, standard tools of quantum optics (master
equations, optical Bloch equations \ldots ) are well adapted to describe
the atom-field interaction for arbitrary intensities of the incident
field, but are restricted to a small number of atoms coupled to each
other by photon exchange \cite{shatokhin05,shatokhin06}. This is due
to the fact that the dimension of the atomic Hilbert space grows
exponentially with the number of atoms.

In the present Letter, we
show that - for the case of two randomly placed atoms with  mean  distance much larger than the laser wavelength -
the two perspectives
of nonlinear
multiple scattering and open system dynamics
can be unified within a single approach.
Our results agree perfectly well with the master
equation and quantum Langevin treatments presented in
\cite{shatokhin07,gremaud06}. In contrast
to \cite{shatokhin07,gremaud06}, however, the present approach can be
generalized
for a large number of atoms.

Imagine a cloud of atoms in free space which is excited by a
coherent laser beam. We are interested in the average intensity
radiated into different directions $\theta$, where the average is
taken over the random
positions of the atoms.  Around the backscattering direction
$\theta=0$,
a coherent
backscattering (CBS) peak is observed \cite{labeyrie99},
originating from the constructive interference of counterpropagating,
multiply scattered waves. In the case of a dilute and optically thin
cloud, the leading contribution to the interference
signal comes from
double scattering events, where the laser interacts only with
pairs of atoms.
(Single scattering does not
contribute to CBS, since
direct and reversed paths
are identical in this case.)
Hereafter, we therefore derive the double scattering
CBS contribution from a simplified model composed of only two
atoms, considered stationary for the time for one scattering cycle to take place, and separated by a distance which is large compared to the laser
wavelength \cite{shatokhin05,shatokhin06}.
For simplicity, we further consider two-level atoms coupled to a scalar photon
field, such as to avoid the additional (though straightforward)
technical overhead needed to account
for photon polarization and degenerate atomic transitions.

\begin{figure}
\centerline{\includegraphics[width=8cm]{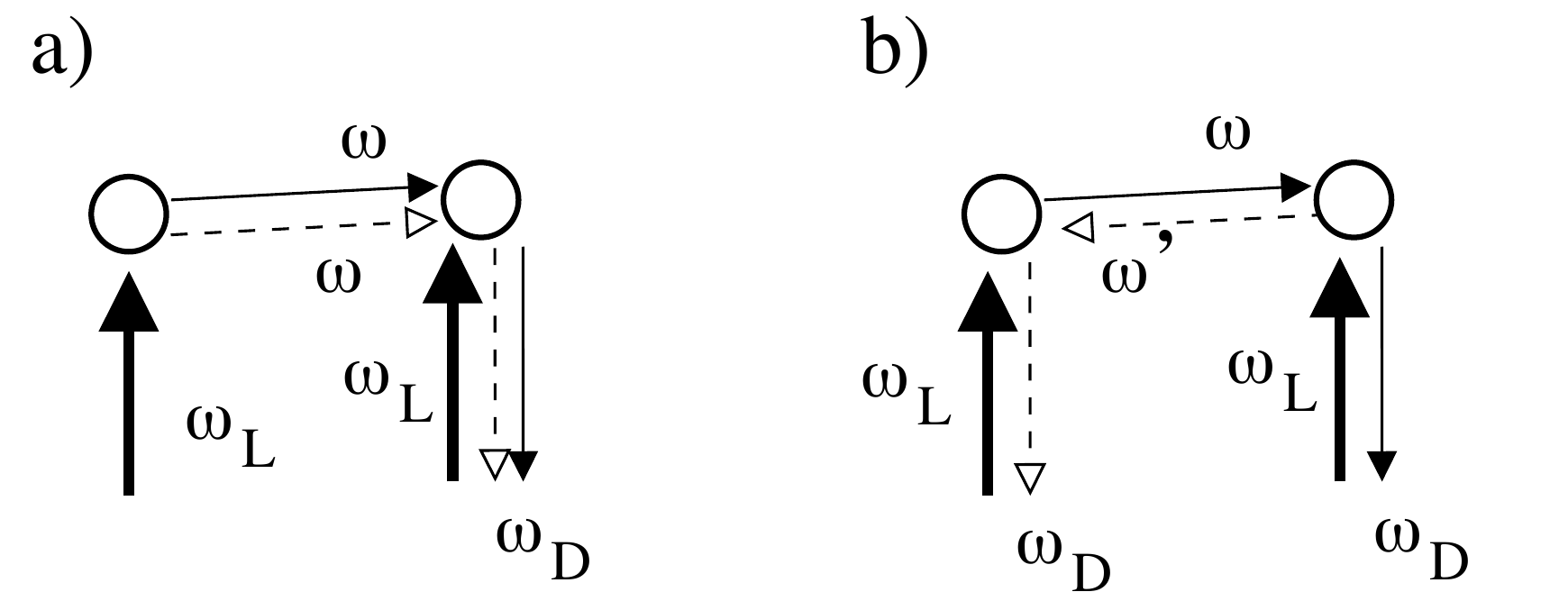}}
\caption{Double scattering contributions to coherent
  backscattering. The thick arrows represent the intense laser beam,
  and the thin solid and dashed arrows the Fourier components of the scattered
  fields, $E^{(+)}(t)$ and $E^{(-)}(t)$, respectively. (a) One of
  the atoms emits a photon with frequency $\omega$ which is scattered by the
  other atom before it reaches the detector. This process yields the
  background intensity $I_L(\omega_D)$ of the photodetection signal. (b) One
  of the
  atoms emits a photon with frequency $\omega$, the other a
  photon with frequency $\omega'=\omega_L+\omega_D-\omega$ (see text).
  This process gives rise to the coherent backscattering
  interference peak $I_C(\omega_D)$.\label{fig1}}
\end{figure}

The situation we consider is summarized in Fig.~\ref{fig1}.
Two atoms scatter photons injected through an intense laser beam at
frequency $\omega_L$ into distinct modes with 
frequencies $\omega$, $\omega^{\prime}$, and $\omega_D$.
While $\omega_D$ is the frequency of the detected photons, $\omega$
and $\omega^{\prime}$ are the frequencies of the (unobserved)
photons exchanged between the atoms -- thus mediating the double
scattering process. Fig.~\ref{fig1}a defines the weakly
angle-dependent background of the photodetection intensity, also
called ``ladder intensity'' $I_L$ in the following, whereas
Fig.~\ref{fig1}b determines the height $I_C$ (``crossed intensity'')
of the coherent backscattering peak around $\theta=0$, resulting
from interference between photons travelling along reversed
scattering paths. Both $I_L$ and $I_C$ are independent of the positions of the two atoms, and thus define, for the case of large mean distance between the atoms, the configuration-averaged photodetection intensity \cite{shatokhin05,shatokhin06}.  The assumption of large distance is required in order to neglect processes where more than one photon is exchanged between the atoms.

The key ingredient for our subsequent treatment is the observation that
the detected double scattering signal can be deduced from the superposition of
the fields scattered by both individual
atoms into the detector, each atom being subject
to a {\em bichromatic} driving field ${\cal E}(t)$ composed of
the injected laser {\em and} the
photons
scattered by the other atom,
\begin{equation} {\cal
E}(t)={\cal E}_Le^{-i\omega_L t} +v^{(+)} e^{-i\omega t}
+{\cal E}_L^*e^{i\omega_L t}+ v^{(-)} e^{i\omega t}\, .\label{drive}
\end{equation}
The classical field ${\cal E}(t)$ is decomposed into its positive and negative frequency
components with complex amplitudes ${\cal E}_L$ and $v^{(+)}$($=v^{(-)*}$),
since
the final figure of merit, the average stationary intensity $\langle
I\rangle=\lim_{t\to\infty}\left<E^{(-)}(t)E^{(+)}(t)\right>$ measured by the
detector, is conveniently expressed in terms of positive and negative frequency
components of the signal radiated by the two atoms.

The single atom response to
the bichromatic drive is
characterized by the associated cross correlation function in
frequency space, $C(\omega_1,\omega_2)=\langle
\tilde{E}^{(-)}(\omega_1)\tilde{E}^{(+)}(\omega_2)\rangle$, which
correlates the, in general, different frequencies of the negative
(thin dashed lines in Fig.~\ref{fig1}) and positive (thin continuous
lines in Fig.~\ref{fig1}) frequency components
$\tilde{E}^{(-)}(\omega_1)$ and  $\tilde{E}^{(+)}(\omega_2)$ of the
emitted field (left atom: $\omega_1=\omega_D$ and $\omega_2=\omega$,
right atom: $\omega_1=\omega'$ and $\omega_2=\omega_D$, in
Fig.~\ref{fig1}b). Since the emitted fields $E^{(-)}$ and $E^{(+)}$,
in turn, are proportional to the atomic dipole raising and lowering
operators $\sigma_+$ and $\sigma_-$, the frequency correlation
function can be written as:
\begin{equation}
C(\omega_1,\omega_2)=\int_{-\infty}^\infty \frac{dt_1dt_2}{2\pi}
  e^{-i\omega_1 t_1+i\omega_2 t_2}
  \langle\sigma_+(t_1)\sigma_-(t_2)\rangle\label{dipole}\, ,
\end{equation}
and is thus given by the solutions $\sigma_+(t)$ and $\sigma_-(t)$
of the {\em single atom} optical Bloch
equations under the bichromatic drive, Eq.~(\ref{drive}).

The total detected double scattering
intensity at frequency $\omega_D$ is finally given by an integral over the
product of the cross correlation functions of both atoms, where the integral
runs over the frequencies of the exchanged photons.
For an explicit
evaluation of these integrals, we expand $C(\omega_1,\omega_2)$ to lowest
order in $v^{(+)}$ and $v^{(-)}$. This perturbative treatment is actually
already implied by our ansatz of {\em classical} scattered fields in
Eq.~(\ref{drive}), which neglects their quantum statistical properties -- encoded
in the second order correlation functions, i.e. in higher orders of $v^{(\pm
  )}$ -- and is certainly justified in the presently considered limit of a
dilute gas with large interatomic distances and, thus, $|v^{(\pm)}|\ll |{\cal
  E}_L|$. Solving Eq.~(\ref{dipole}) for the case of bichromatic driving, Eq.~(\ref{drive}), and keeping only terms
up to first order in $v^{(+)}$ and/or $v^{(-)}$, we obtain
the following four terms:
\begin{subequations}
\begin{eqnarray}
\left.C\right|_{v^{(\pm)}=0}
& = & \delta(\omega_1-\omega_2)P_0(\omega_1)\label{p0},\\
\left.\frac{\partial C}{\partial v^{(+)}}\right|_{v^{(\pm)}=0} & = &
\delta(\omega_2-\omega_1-\omega+\omega_L)
P_1(\omega;\omega_2),\ \ \ \ \ \label{c2} \\
\left.\frac{\partial C}{\partial v^{(-)}}\right|_{v^{(\pm)}=0}
& = & \delta(\omega_1-\omega_2-\omega+\omega_L)P^*_1(\omega;\omega_1)\label{c1},\\
\left.\frac{\partial^2 C}{\partial v^{(+)}\partial
    v^{(-)}}\right|_{v^{(\pm)}=0} &
= & \delta(\omega_1-\omega_2)P_2(\omega;\omega_1)\label{p1}.
\end{eqnarray}
\end{subequations}
The $\delta$-functions in Eqs.~(\ref{p0}-\ref{p1})
originate from integrating over $t_+:=t_1+t_2$
in Eq.~(\ref{dipole}), and are thus a consequence
of time translation invariance or energy conservation.
They ensure that the negative and positive frequency components of the
field are
radiated at the same frequency $\omega$ in Fig.~\ref{fig1}a, whereas
$\omega'=\omega_L+\omega_D-\omega$ in Fig.~\ref{fig1}b.

$P_0(\omega)$ in Eq.~(\ref{p0}) denotes the usual resonance
fluorescence spectrum for monochromatic driving including both the
elastic and inelastic components \cite{mollow69}, and is
associated with the emission of a fluorescence photon of frequency
$\omega$ by the left atom in Fig.~\ref{fig1}a.
$P_2(\omega;\omega_D)$, defined by Eq.~(\ref{p1}), describes the
subsequent scattering cross section of this fluorescence photon by
the right atom, into the finally detected frequency mode $\omega_D$.
Consequently, the final expression for the ladder contribution
(Fig.~\ref{fig1}a) to the total scattered intensity at frequency
$\omega_D$ reads
\begin{equation}
I_L(\omega_D)  =  |g|^2\int d\omega
P_0(\omega)P_2(\omega;\omega_D)\, ,\label{ilad}
\end{equation}
where $g=\exp(i\omega_L r)/(k_Lr)$, with $k_L=\omega_L/c$,
  describes the effective coupling
strength between the two atoms. The frequency
dependence of $g$ can be neglected for inverse propagation times
$c/r$ between the two atoms much larger than the frequency
differences $\omega'-\omega$ between the exchanged photons (which is
the case for typical experimental parameters). Note that the
integral over the frequency $\omega$ of the photon emitted by the
left atom expresses the correlation
between
both atoms induced through the photon exchange.

Analogously, according to Eq.~(\ref{c1}), $P_1^*(\omega';\omega_D)$ 
represents the emission of a negative frequency {\em amplitude}
$\omega_D$ (incident on the detector) and a positive frequency {\em
amplitude} $\omega$ (towards the other atom) by the left atom in
Fig.~\ref{fig1}b, which, in turn, is subject to a negative frequency
amplitude $\omega'$ originating from the right atom. The right atom,
subject to a positive frequency amplitude $\omega$ from the left
atom, radiates a negative frequency amplitude $\omega^\prime$
(towards the left atom) and a positive frequency amplitude
$\omega_D$ (into the detector), represented by
$P_1(\omega;\omega_D)$ in Eq.~(\ref{c2}). Energy conservation
further enforces $\omega^\prime =\omega_L+\omega_D-\omega$.
For the crossed contribution to the total scattered intensity at the
frequency $\omega_D$, this entails
\begin{equation}
I_C(\omega_D) = |g|^2\int d\omega
P^*_1(\omega_L+\omega_D-\omega;\omega_D)P_1(\omega;\omega_D)\, .\label{icr}
\end{equation}
Once again, both atomic dipoles are conditioned on each other
through the frequency of the exchanged photon. Finally, the total
intensities measured by a broadband detector are obtained by
integrating over the detected photon frequency, i.e., $I_{L,C}=\int
d\omega_D I_{L,C}(\omega_D)$.

\begin{figure}
\centerline{\includegraphics[width=8cm]{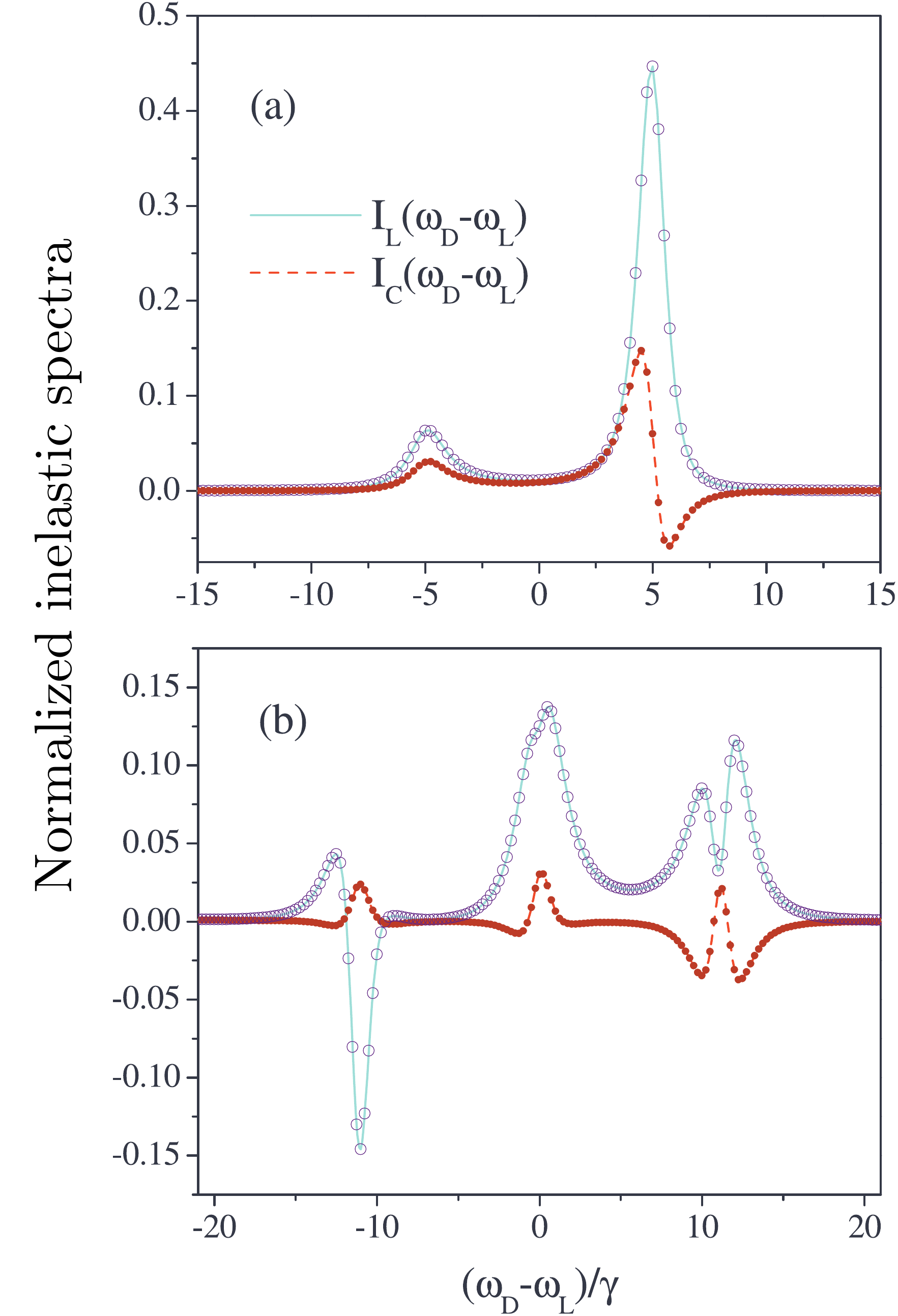}}
\caption{(Color online) Inelastic double scattering
spectra as obtained from the two-atom master equation (solid line: background, dashed line: interference contribution),
compared to the predictions of our new approach, Eqs.~(\ref{ilad},\ref{icr}), based on single-atom Bloch equations with bichromatic driving 
(open circles: background, filled circles: interference contribution). Laser-atom detuning $\delta=-5\gamma$, 
Rabi frequency $\Omega=$ (a) $0.1\gamma$ and (b) $10\gamma$. Perfect agreement is observed, which highlights the exact equivalence of both approaches.\label{spectra}}
\end{figure}

By numerical integration of Eqs. (\ref{ilad}) and (\ref{icr}), we have
studied the behavior of the elastic and inelastic components of
the ladder and crossed terms for different values of the laser-atom
detuning $\delta=\omega_L-\omega_0$ (with $\omega_0$ the atomic resonance frequency) and of the Rabi frequency $\Omega$ (which is proportional to the laser amplitude $|{\mathcal E}_L|$). In particular,
some examples of the normalized {\em inelastic} ladder and crossed
spectra for several parameters of the driving field are presented in
Fig.~\ref{spectra}.
At small Rabi frequencies (see Fig. \ref{spectra}(a)), the background and interference spectra are
qualitatively very similar to the analogous quantities in the helicity preserving polarization channel \cite{wellens04,shatokhin07}, and can be deduced from the two-photon scattering amplitudes.
When $\Omega\gg \gamma$ (Fig. \ref{spectra}(b)), the spectral features
of the ladder and crossed terms can be interpreted in terms of scattering and
self-interference of photons emitted by one atom on the dressed
state levels of the other atom \cite{shatokhin07}. Note that negative values of the ladder term
in certain spectral regions arises from absorption dominating emission for doubly scattered photons;
the total contribution, including single scattering, is always positive.

For comparison with these results for the single-atom response under bichromatic driving, we calculate the same quantities
-- the ladder and crossed double scattering spectra -- using the
familiar two-atom master equation approach
\cite{shatokhin05,shatokhin06,shatokhin07}. As already mentioned,
this approach provides a general framework for treating intense
laser field-atoms interactions including interatomic correlations
induced by the exchange of photons, but is restricted to small
numbers of scatterers. 
 Both, $I_{L}(\omega_D)$ and $I_{C}(\omega_D)$, derived from the two-atom master
 equation, are found to coincide  perfectly  with the results of
 Eqs.~(\ref{ilad},\ref{icr}), for all tested sets of the driving
 field parameters, see Fig.~\ref{spectra}. Motivated by this agreement between the
 two apparently very different approaches, we succeeded to prove their equivalence analytically - at least for the elastic component of the scattered light \cite{diplgeiger}. We expect that the proof can be extended to the inelastic component, as suggested by the good numerical agreement observed in Fig.~\ref{spectra}. Details of the proof
will be given elsewhere.

Let us stress again that the reduction of the two-atom problem to single-atom Bloch equations is possible only in the case of large distance $k_Lr\gg 1$ (and hence $|g|\ll 1$) between the atoms. In this regime, both atoms exchange only single photons, what, in turn, allows to describe the field acting on each atom by a classical field, see Eq.~(\ref{drive}). The same argument remains valid for a multiple scattering process
with more than two atoms. Provided that the atoms are far away from
each other, there are again only single photons exchanged between
the individual atoms. Moreover, the possibility that two photons emitted by the same atom meet again at another atom before leaving the sample can be neglected. In other words, all photons incident on a particular atom originate from different previous atomic scatterers, and are hence
uncorrelated with each other. Since the quantum properties of the electromagnetic field manifest themselves only in correlations between different photons, the field inside the disordered sample can therefore be described by a classical field. This allows, in principle, to generalize the diagrammatic
multiple scattering theory valid for classical nonlinear scatterers
\cite{wellens07} to the atomic case. The main difficulty to be
addressed in future work will be to find an efficient method to
solve the {\em single atom} optical Bloch equation for a
fluctuating, polychromatic field, representing the radiation emitted
by many randomly placed atoms.

In summary, we have shown that the spectrum of intense laser light
scattered by two distant atoms can be reproduced by solving the quantum
mechanical
time evolution equations for a {\em single} atom in the presence of
a weak probe field representing the light emitted by the
other atom. The interaction between the atoms is fully taken into
account by integrating over the frequencies of the exchanged
photons. Since, in contrast to other standard methods in theoretical quantum
optics, it is {\em not} necessary to perform calculations in the composite
Hilbert
space of both atoms, our method will exhibit very favourable scaling
with increasing system size, and can be applied
to a larger number
of atoms. Thereby, it can
serve as a starting point for the
development of a theory of
multiple scattering of intense laser light by optically thick samples
of cold atoms.

\end{document}